\documentclass[preprint2]{aastex}
\usepackage{natbib}
\usepackage{lscape}
\usepackage{wasysym}
\usepackage{txfonts}
\usepackage{graphicx}
\usepackage{hyperref}
\bibpunct{(}{)}{;}{a}{}{,}
\usepackage{longtable}
\usepackage{hyperref}

\usepackage{afterpage}

\usepackage[usenames,dvipsnames]{color}
\definecolor{mygreen}{rgb}{0,0.5,0}

\def\ms{\hbox{\,m\,s$^{-1}$\,}}         %m.s -1
\def\cms{\hbox{\,cm\,s$^{-1}$\,}}       %cm.s -1
\def\m2s2{\hbox{\,m$^{2}$\,s$^{-2}$}\,} %m2.s -2
\def\kms{\hbox{\,km\,s$^{-1}$}\,}       %km.s -1
\newcommand{\titlestar}{\star}

\shorttitle{HARPS-N observes the Sun as a star}
\shortauthors{X. Dumusque}

\begin{document}

\title{HARPS-N observes the Sun as a star}%\altaffilmark{\titleast}}

\author{Xavier Dumusque\altaffilmark{1}\altaffilmark{\titlestar}
		, Alex Glenday\altaffilmark{1}
		, David F. Phillips\altaffilmark{1}
		, Nicolas Buchschacher\altaffilmark{2}
		, Andrew Collier Cameron\altaffilmark{3}
		, Massimo Cecconi\altaffilmark{4}
		, David Charbonneau\altaffilmark{1}
		, Rosario Cosentino\altaffilmark{4}
		, Adriano Ghedina\altaffilmark{4}
		, David W. Latham\altaffilmark{1}
		, Chih-Hao Li\altaffilmark{1}
		, Marcello Lodi\altaffilmark{4}
		, Christophe Lovis\altaffilmark{2}
		, Emilio Molinari\altaffilmark{4,5}
		, Francesco Pepe\altaffilmark{2}
		, St\'ephane Udry\altaffilmark{2}
		, Dimitar Sasselov\altaffilmark{1}
		, Andrew Szentgyorgyi\altaffilmark{1}
		, Ronald Walsworth\altaffilmark{1}}

\altaffiltext{1}{Harvard-Smithsonian Center for Astrophysics, 60 Garden Street, Cambridge, Massachusetts 02138, USA}
\altaffiltext{2}{Observatoire Astronomique de l'Universit\'e de Gen\`eve, 51 Chemin des Maillettes, 1290 Sauverny, Suisse}
\altaffiltext{3}{SUPA, School of Physics \& Astronomy, University of St. Andrews, North Haugh, St. Andrews Fife,
KY16 9SS, UK}
\altaffiltext{4}{INAF - Fundaci\'o—n Galileo Galilei, Rambla Jos\'eŽ Ana Fern\'andez P\'eŽrez 7, 38712 Bre$\tilde{–\mathrm{n}}$a Baja, Spain}
\altaffiltext{5}{INAF - IASF Milano, via Bassini 15, 20133, Milano, Italy}

\altaffiltext{$\star$}
{xdumusque@cfa.harvard.edu}

%\altaffiltext{$\ast$}
%{Based on observations made with the HARPS instrument on the ESO 3.6-m telescope at La Silla Observatory (Chile).}

\begin{abstract}
% Context, Aims, Metods, Results, Conclu (not mandatory)
Radial velocity perturbations induced by stellar surface inhomogeneities including spots, plages and granules currently limit the detection of Earth-twins using Doppler spectroscopy. Such stellar noise is poorly understood for stars other than the Sun because their surface is unresolved. In particular, the effects of stellar surface inhomogeneities on observed stellar radial velocities are extremely difficult to characterize, and thus developing optimal correction techniques to extract true stellar radial velocities is extremely challenging. In this paper, we present preliminary results of a solar telescope built to feed full-disk sunlight into the HARPS-N spectrograph, which is in turn calibrated with an astro-comb. This setup enables long-term observation of the Sun as a star with state-of-the-art sensitivity to radial velocity changes. Over seven days of observing in 2014, we show an average 50\cms radial velocity rms over a few hours of observation. After correcting observed radial velocities for spot and plage perturbations using full-disk photometry of the Sun, we lower by a factor of two the weekly radial velocity rms to 60\cms. The solar telescope is now entering routine operation, and will observe the Sun every clear day for several hours. We will use these radial velocities combined with data from solar satellites to improve our understanding of stellar noise and develop optimal correction methods. If successful, these new methods should enable the detection of Venus over the next two to three years, thus demonstrating the possibility of detecting Earth-twins around other solar-like stars using the radial velocity technique.
\end{abstract}


\section{Introduction} \label{sect:1}

The radial velocity (RV) technique is successful at finding super-Earth planets orbiting bright stars \citep[][]{Mayor-2011,Howard-2010}, as well as extremely short period Earth-mass planets \citep[][]{Dressing-2015,Dumusque-2014a,Pepe-2013, Howard-2013b,Dumusque-2012}. However, RV detection of Earth-twins, i.e., long-period terrestrial worlds around solar-type stars, requires an order of magnitude improvement in precision to $\sim$10\cms. Among the multitude challenges to reaching such RV precision, the most important ones are: (i) stable, long-term wavelength calibration of the astrophysical spectrograph; and (ii) accounting for the effects of stellar noise, i.e., RV perturbations induced by stellar surface inhomogeneities, which are typically a few \ms for solar-type stars. The first challenge has been successfully addressed with astro-comb calibrators, based on laser frequency combs locked to atomic clocks \citep[][]{Glenday-2015,Phillips-2012b,Wilken-2012,Wilken-2010}. To meet the second challenge, it will be crucial to better understand RV noise induced by stars and develop mitigation techniques, particularly as current and future high-resolution spectrographs such as HARPS \citep[with its newly commissionned astro-comb,][]{Wilken-2012,Mayor-2003}, HARPS-N \citep[][]{Cosentino-2012}, ESPRESSO \citep[][]{Pepe-2014}, and G-CLEF \citep[][]{Szentgyorgyi-2014} should reach long-term instrumental RV precision allowing Earth-twin detection.

Stellar RV noise can be differentiated into four categories (given the current state of knowledge) : 
\begin{itemize}
\item stellar oscillations, on a timescale of a few minutes for solar-like stars, produced by pressure waves propagating in stellar interiors \citep[][]{Dumusque-2011a,Arentoft-2008,Kjeldsen-2005};
\item stellar granulation and supergranulation, on a timescale from a few minutes to 48 hours, resulting from convectively driven outflowing material reaching the stellar surface \citep[][]{Dumusque-2011a,Del-Moro-2004a,Del-Moro-2004b};
\item short-term stellar activity, on the timescale of the stellar rotational period, induced by rotation in the presence of evolving and decaying surface inhomogeneities including spots and plages \citep[e.g.,][]{Borgniet-2015,Dumusque-2014b,Boisse-2012b,Saar-2009,Meunier-2010a,Saar-1997b}; and
\item long-term stellar activity, on a timescale of several years, induced by stellar magnetic cycles and/or surface flows \citep[][]{Meunier-2013,Lovis-2011b,Dumusque-2011c,Makarov-2010}.
\end{itemize}
Due to their semi-periodic nature, stellar oscillations can be averaged out efficiently by taking measurements with exposure times longer than the timescale of these perturbations \citep[][]{Dumusque-2011a}. This observational strategy is now commonly employed in high precision RV surveys, e.g., by imposing a minimum exposure time of fifteen minutes. Stellar granulation can be mitigated somewhat by observing a star several times per night, with the maximum possible spread between measurements \citep[][]{Dumusque-2011a}. However, this approach is only partially effective, and better mitigation methods need to be developed. For short-term activity, some correction techniques have been investigated \citep[][]{Meunier-2013,Aigrain-2012,Boisse-2011,Boisse-2009}, but none are fully satisfactory. Finally, long-term activity seems to correlate well with the calcium chromospheric activity index, which provides a promising approach to mitigation of this source of stellar RV noise \citep[][]{Meunier-2013,Dumusque-2012}.

Characterizing the effects of stellar noise sources on stellar spectra with sufficient precision to allow the detection of true Earth twins from RV measurements is a challenging task. An orbiting planet will produce on the stellar spectrum a pure shift of all the spectral lines, while stellar noise will modify the shape of spectral lines. The two effects are different in nature and should be distinguishable. However, when considering an Earth-like planet orbiting a quiet star like the Sun, both the planetary and stellar signals will induce shifts or shape variations of spectral line smaller than a thousand of a detector pixel. Therefore, we cannot distinguish one effect from the other on single spectral lines. To get around the problem, we need to average thousands of spectral lines to construct observables that are sensitive to stellar noise. However this is extremely challenging as this noise affects each spectral line differently.

\begin{figure*}[t]
\begin{center}
\includegraphics[width=16cm]{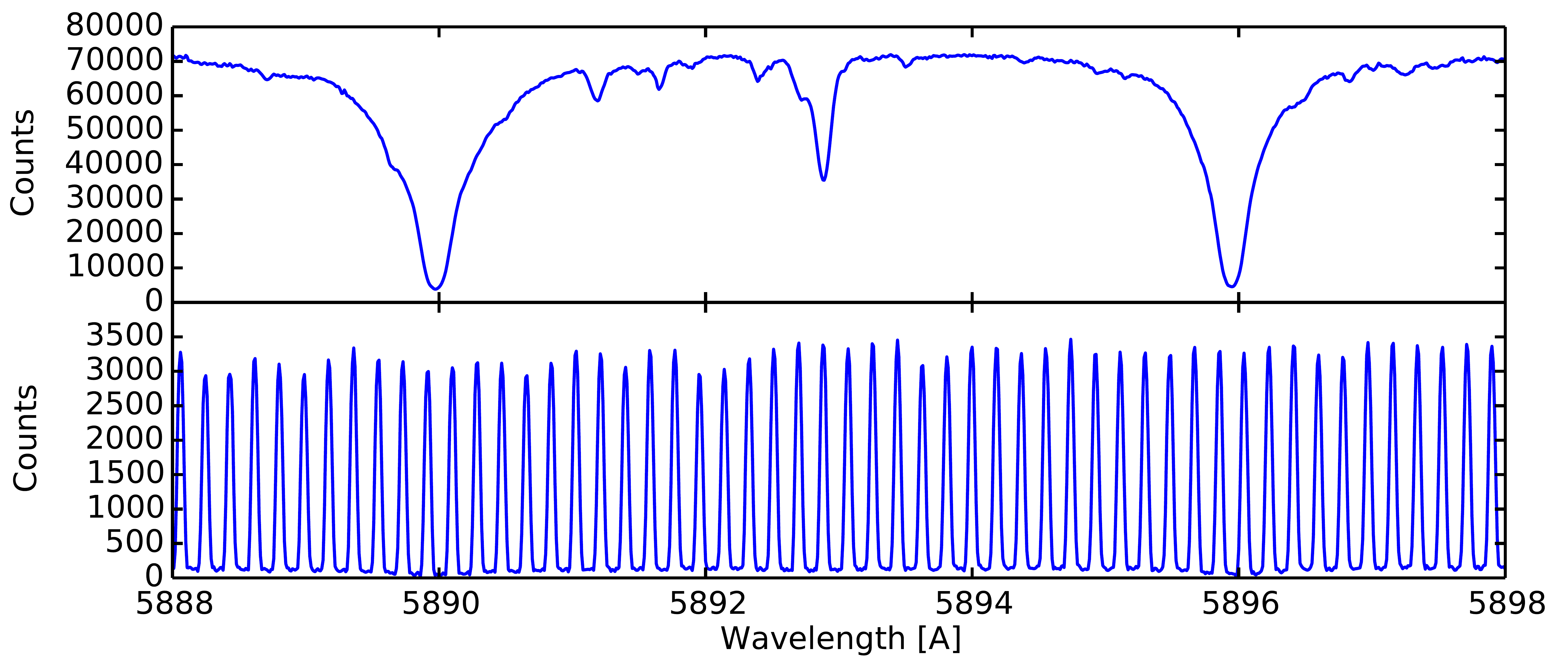}
\caption{High signal-to-noise solar spectrum centered on the Sodium lines (top) with its simultaneous astro-comb wavelength calibration (bottom) as observed by HARPS-N.}
\label{fig:3-10}
\end{center}
\end{figure*}

So far, it was only possible to obtain precise, frequent, full-disk RV measurements derived from the full optical spectrum for stars other than the Sun. However, the surfaces of other stars are not resolvable; and therefore all the information used to study surface inhomogeneities is indirect, including: photometric flux \citep[e.g.,][]{Aigrain-2012,Boisse-2009}, bisector of spectral lines or of the cross correlation function depending on the activity level of the star \citep[e.g.,][]{Dumusque-2014b,Figueira-2013,Queloz-2001,Vogt-1987}, or the calcium chromospheric activity index \citep[][]{Noyes-1984}. Because it is extremely difficult to infer from such indirect observables the size, location, and contrast of surface inhomogeneities, understanding in detail the RV perturbations produced by these features is a major challenge.

As a way forward, we note that we might better understand the effect of surface inhomogeneities on RVs by obtaining precise full-disk RV measurements of the Sun. This approach of observing the Sun as a star allows us to directly correlate any change in surface inhomogeneities observed by solar satellites like the Solar Dynamics Observatory \citep[SDO,][]{Pesnell-2012} with variations in the full-disk RV. Thus, we built a small solar telescope to feed a full-disk unresolved image of the Sun into the HARPS-N spectrograph located at the Telescopio Nazionale Galileo at the Observatorio del Roque de los Muchachos, Spain. We plan to use this instrument to observe the Sun several hours every clear day over the next two to three years, employing the exquisite long-term sub\,-\ms precision of HARPS-N calibrated by an astro-comb. Note that this project is not the only one to feed full-disk sunlight into a high-precision spectrograph \citep[][]{Strassmeier-2015,Probst-2015,Kjeldsen-2008}. However, our setup observes the Sun using exactly the same instrumental configuration as for other HARPS-N targets, which allows for a direct comparison between solar and stellar spectra.
%For more information about these projects, readers are referred to \citet{Probst-2015} for a solar telescope at the Solar Vacuum Tower Telescope on Tenerife, and \citet{Strassmeier-2015} for a solar telescope feeding the PEPSI spectrograph on the Large Binocular Telescope located at the Mount Graham observatory.

%In this paper, we first present briefly in Section \ref{sect:2} the instrumental setup that allows us to get a full-disk unresolved image of the Sun. Section \ref{sect:3} is dedicated to the analysis of the first test measurements, and Section \ref{sect:4} concludes the paper by presenting future milestones of the instrument.

\section{The HARPS-N solar telescope} \label{sect:2} 

\begin{figure*}[t]
\begin{center}
\includegraphics[width=8cm]{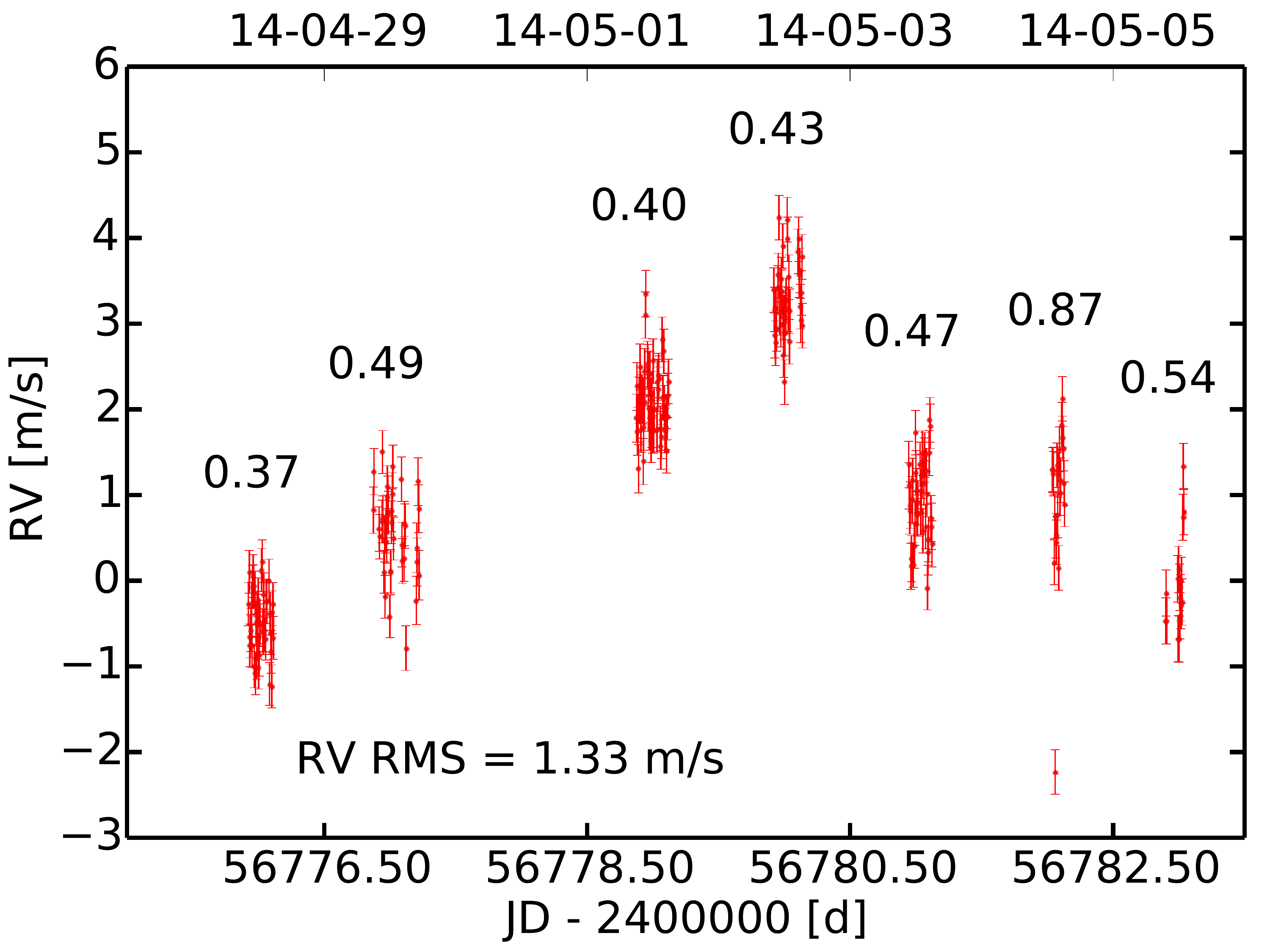}
\includegraphics[width=8cm]{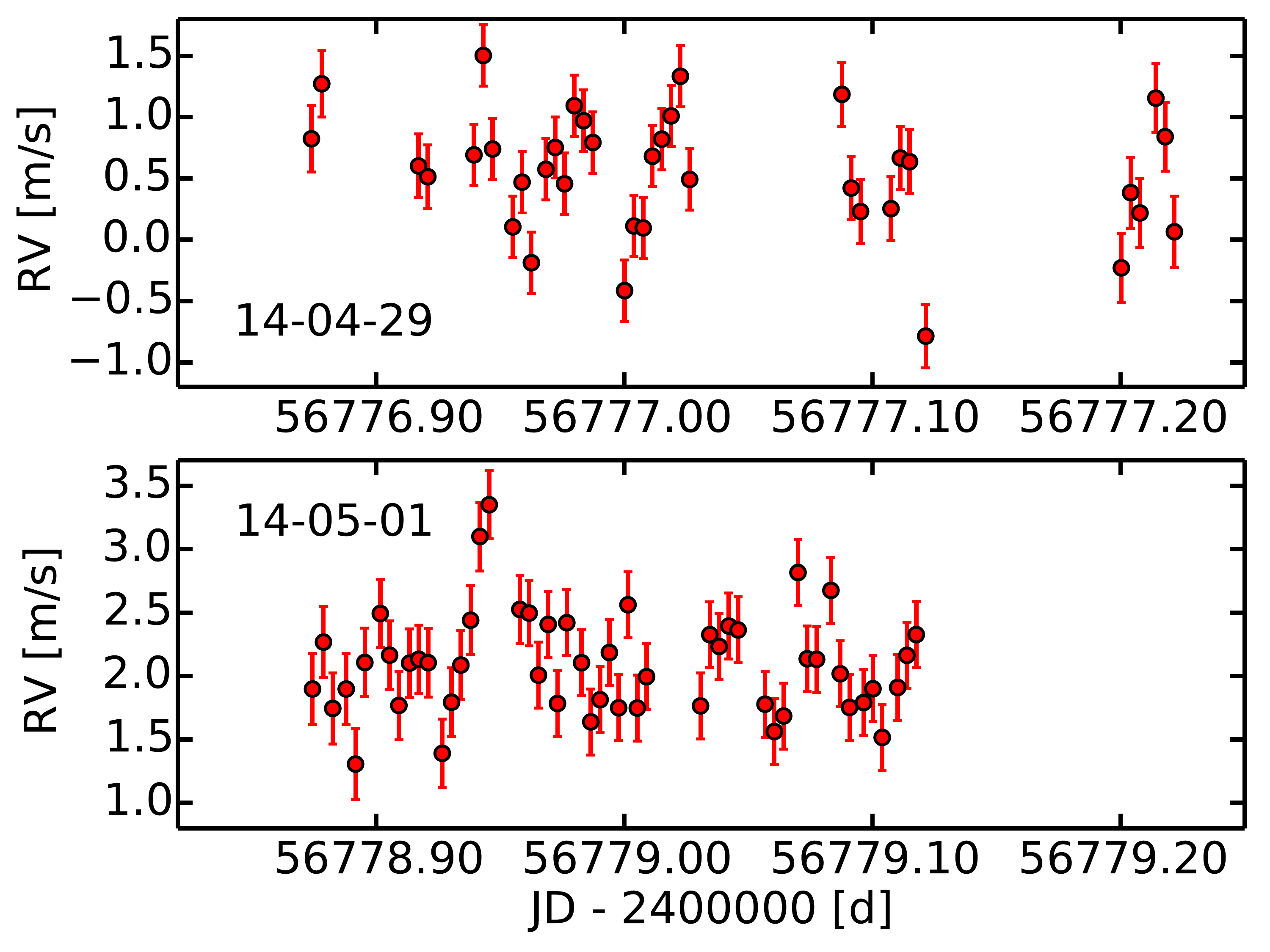}
\caption{\emph{Left:} Solar radial velocities (RVs) after barycentric correction obtained by the HARPS-N solar telescope during one week using simultaneous astro-comb wavelength calibration. The RV rms over the week time span of the observation is 133\cms, and the daily RV rms is shown above each daily observations on the plot. \emph{Right:} Subset of RVs highlighting the daily variations seen on 2014 April 29 and 2014 May 01.}
%
% \caption{\emph{Top: }Raw RVs obtained by the HARPS-N solar telescope. The red points are measurements obtained with simultaneous laser frequency comb (LFC) calibration source. Blue dots are obtained with the standard Thorium lamp calibration source. \emph{Bottom: }Zoom on the RVs obtain for each of the three observing runs. The daily rms is shown on top of the measurements of each observing day. The total rms of each run is also shown.}
\label{fig:3-0}
\end{center}
\end{figure*}

The solar telescope is a simple instrument built from off-the-shelf components. Details of this instrument and its application to solar observations will be provided in a future publication (Glenday et al., in prep.). Here we provide a brief summary. The solar telescope consists of a $3^{\prime\prime}$ $f=200$ mm achromatic lens, which feeds an integrating sphere to scramble the solar disk. A guide camera mounted on the telescope frame provides sufficient guiding to keep the 2 mm solar image centered on a 6 mm input aperture. These components are mounted on an amateur telescope tracking mount following the Sun. The integrating sphere couples to a 300 micron core optical fiber which is fed to the HARPS-N spectrograph via its calibration unit. Light from the Sun is coupled out of the integrating sphere to HARPS-N with a uniformity of better than $10^{-4}$ as measured both in the laboratory and on-sky. This good uniformity is crucial to average out efficiently the 4 \kms RV shift that can be seen between the blueshifted approaching limb and the redshifted receding limb of the Sun. Given these 4 \kms, a uniformity better than $10^{-4}$ enables a quantitative comparison of disk-integrated RV below the 10\cms level.

The throughput of the solar telescope allows for exposures as short as 20 seconds. As described above, to reduce the effects of the solar 5-minute oscillations, we typically expose for five minutes to average over one such oscillation period. However, this instrument can also be used in its 20 second exposure time mode ($\approx45$ seconds including readout) where the five minute oscillation signal is easily observable (Glenday et al., in prep.). Such a mode may prove useful for asteroseismology studies, as well as testing the effectiveness of different exposure times and observing strategies for averaging out oscillation and granulation jitter.
\begin{figure}[!h]
\begin{center}
\includegraphics[width=8cm]{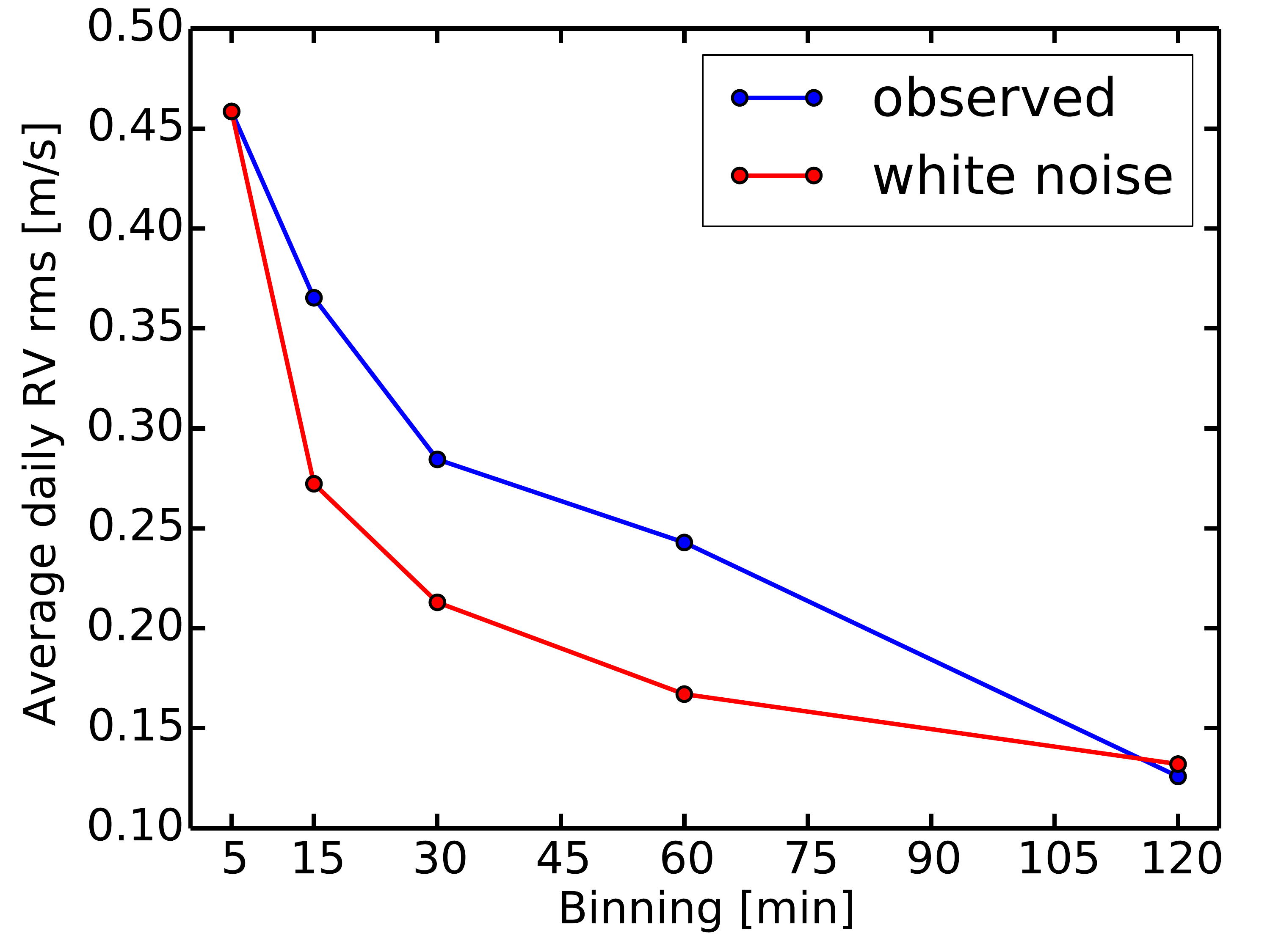}
\caption{Average daily RV rms (e.g., for 5 minutes the average of the values shown on the left of Figure 2) as a function of binning consecutive RV observations. The outlier on 2014 May 04 was not considered for this analysis. Note that for 120 minutes and larger binning, we do not have sufficient observations per day to give a reliable estimates of the average daily RV rms, which explains why the observed RV rms for 120 minutes is smaller than what is expected in the ideal case of white noise.}
\label{fig:3-1}
\end{center}
\end{figure}

\section{First test results} \label{sect:3}

\subsection{Raw solar RVs} \label{sect:3:1}

We first observed the Sun for seven days between April and May 2014 using the simultaneous astro-comb wavelength calibration of HARPS-N to provide RV precision better than 10\cms for the instrumental drift \citep[][]{Glenday-2015,Phillips-2012b}. We selected an exposure time of five minutes for each observation to average out the signal induced by solar oscillations. Figure \ref{fig:3-10} displays one example of such an observation to highlight the high signal-to-noise solar spectrum obtained with its simultaneous astro-comb wavelength calibration ranging from 5000 to 6200 \AA.

We derived the final RVs using a single wavelength solution and corrected for: i) day-to-day instrumental drift using daily Thorium-Argon calibrations; and ii) same day variations using the simultaneous reference fiber containing the astro-comb calibration. We show in Figure \ref{fig:3-0} the raw RVs obtained from the solar telescope only considering clear sky measurements. The presence of clouds in front of the solar disc will break the flux balance between the blueshifted approaching limb and the redshifted receding limb of the Sun and can easily create several dozens of \ms variations.
%Without considering the 4 days affected by outliers (days with a rms higher than 8 \cms) , we see that 
The daily RV rms of our observations span a range between 37 and 87\cms; however the day for which the rms is 87\cms is affected by an outlier. Over the week time span of this observing run, the RV rms increases to 133 \cms. This RV drift could be explained by surface inhomogeneities evolving and moving on the solar disk because of the 25-day rotational period of the Sun. 

The RVs for 2014 April 29 and 2014 May 01, as shown in the right plot of Figure \ref{fig:3-0}, are affected by variations that exceed errors from photon noise. It is likely that this excess jitter is induced by long-period oscillation modes, and longer timescale stellar signals like granulation and supergranulation. However, it is too early to conclude at this point with so few measurements if this excess daily jitter is indeed induced by the Sun. Binning the measurements over timescales longer than 5 minutes, we observe that the average daily RV rms decreases, a sign of residual signals (see Figure \ref{fig:3-1}). Stellar signals are known to be sources of red noise, which could explain the smaller decrease in RV rms with binning than what is expected for white noise. These issues will be studied in much greater detail once substantially larger numbers of observations have been collected.

\subsection{Correction for RV drift induced by surface inhomogeneities} \label{sect:3:2}

We consulted the contemporaneous SDO images to check if the observed RV drift was caused by inhomogeneities present on the solar surface. In Figure \ref{fig:3-1}, we show the SDO images of the Sun taken on BJD = 2456780 (2 May 2014, 12:00 UT), corresponding to the maximum RV of the first observing run. It is clear from the Helioseismic and Magnetic Imager \citep[HMI][]{Schou-2012} intensity map and magnetogram that large surface inhomogeneities were present on the Sun at this time.
\begin{figure*}[t]
\begin{center}
\includegraphics[width=4cm]{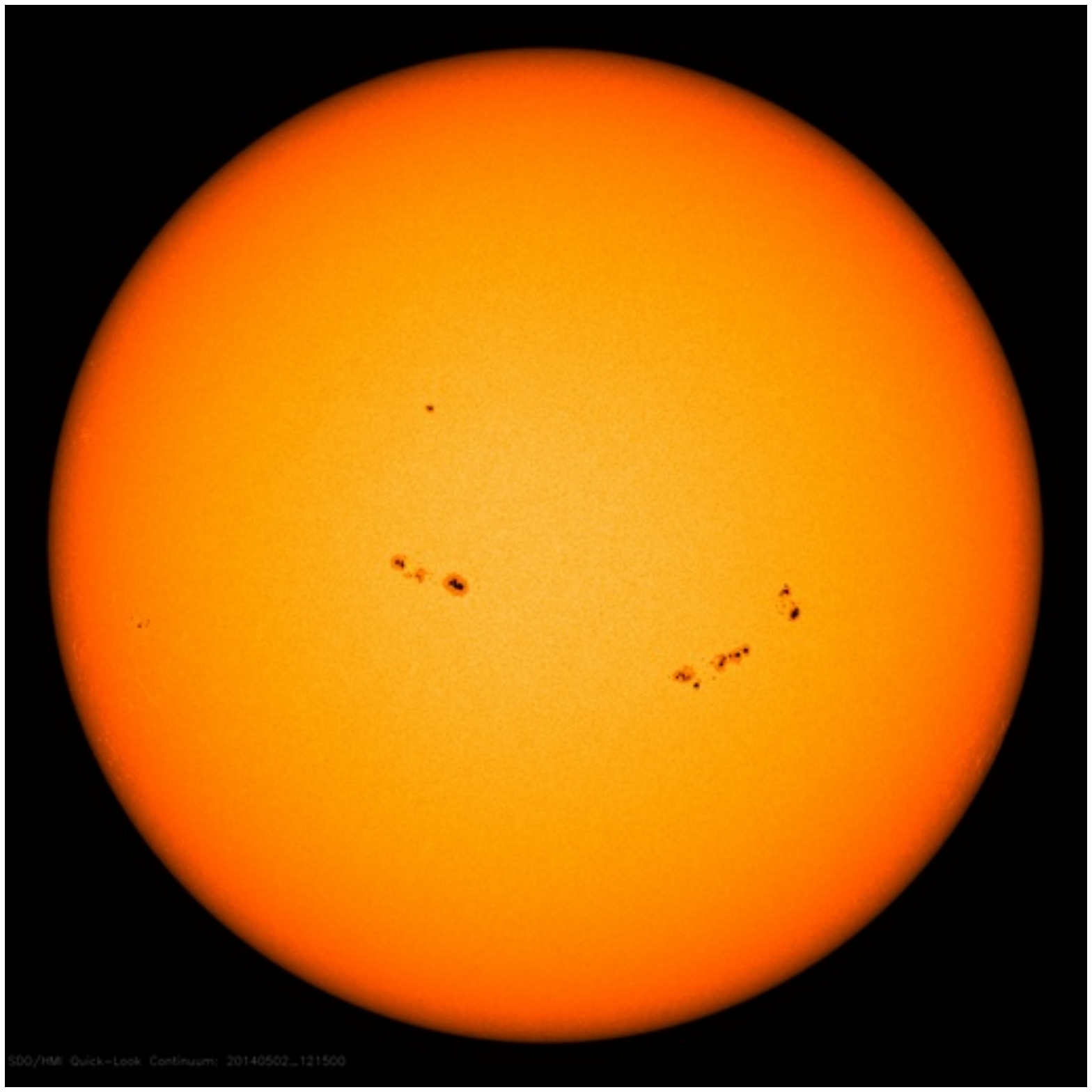} %Data SDO HMI intensitygram color 2 May 2014 between 12:00 and 12:30
\includegraphics[width=4cm]{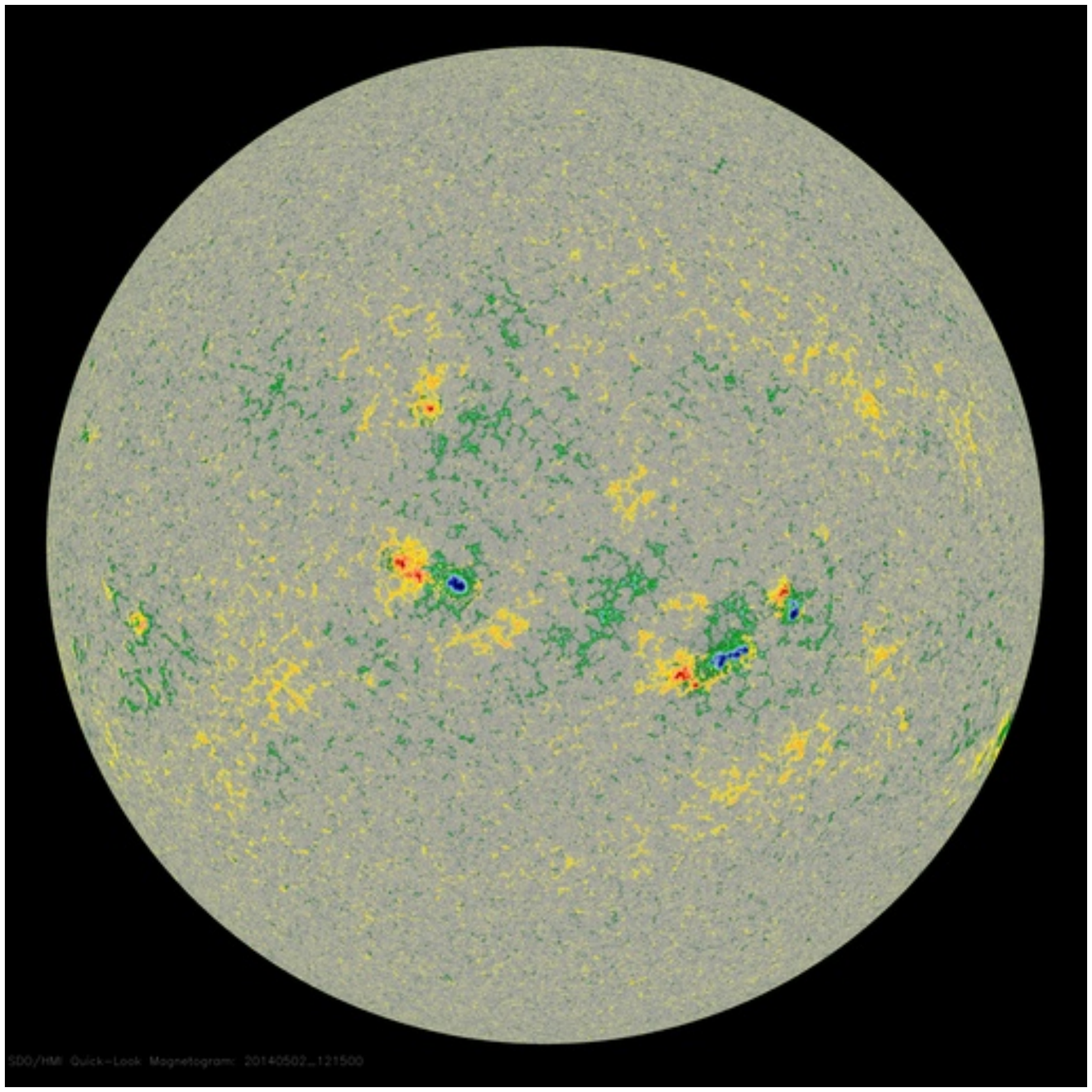} %Data SDO Magnetogram 2 May 2014 between 12:00 and 12:30
\includegraphics[width=4cm]{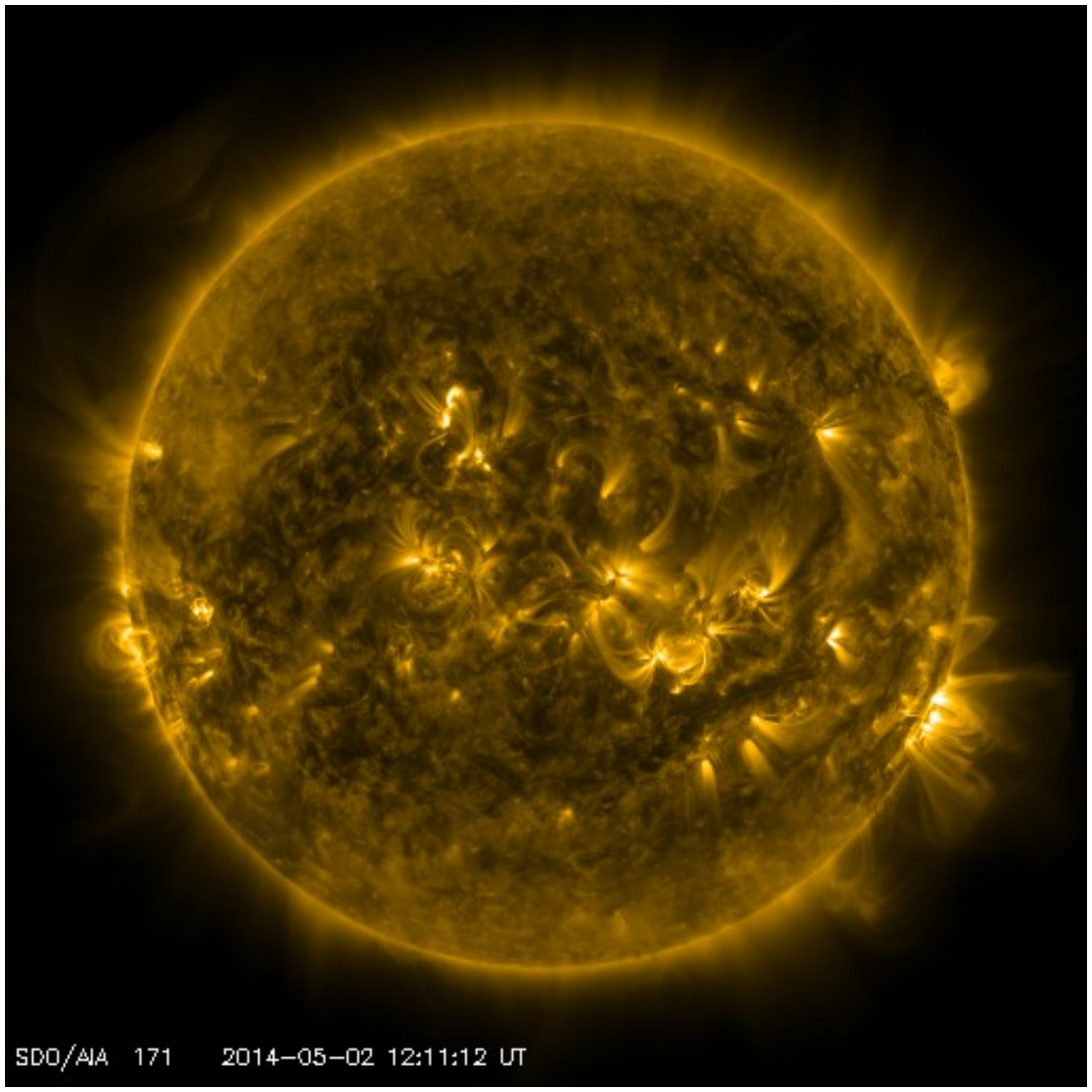} %Data SDO AIA171 2 May 2014 between 12:00 and 12:30
\includegraphics[width=4cm]{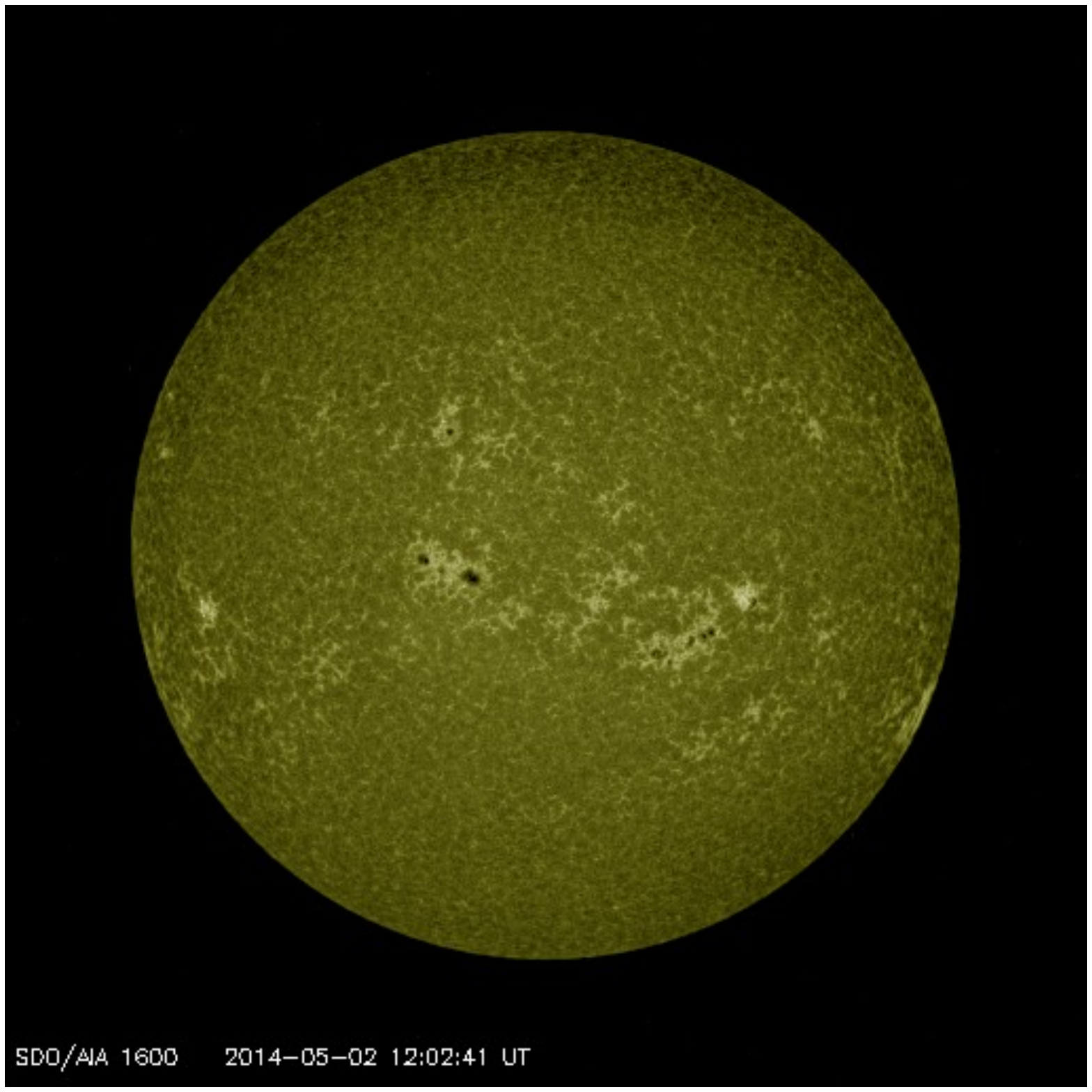} %Data SDO AIA1600 2 May 2014 between 12:00 and 12:30
\caption{SDO images taken on BJD = 2456780 (2 May 2014, 12:00 UT), which corresponds to the maximum RV of the first observation run. From left to right, it is possible to see an HMI intensity map of the Sun, a magnetogram (6173 \AA), and a view of the Sun in extreme ultraviolet (171 \AA) and in the far ultraviolet (1600 \AA). It is clear that the Sun is active at this given time.}
\label{fig:3-1}
\end{center}
\end{figure*}

For a first order correction of the RV drift due to these surface inhomogeneities, we applied the FF${}^\prime$ method proposed by \citet{Aigrain-2012}. These authors show that a simple model estimating the RV effect of spots and plages can link the RV variation to the flux variation using the following formula:
\begin{equation}
\Delta RV = a\times \mathrm{F}\mathrm{F'} + b\times \mathrm{F}^2,
\end{equation}
where F and F${}^\prime$ correspond to the stellar flux and its time derivative. Parameters $a$ and $b$ are proportional to the relative flux drop for a spot at the disk centre, the convective blueshift difference in the unspotted photosphere and that within the magnetized area, and the ratio of the area covered by plages to the one covered by spots. 

\begin{figure}
 \begin{center}
 \includegraphics[width=8cm]{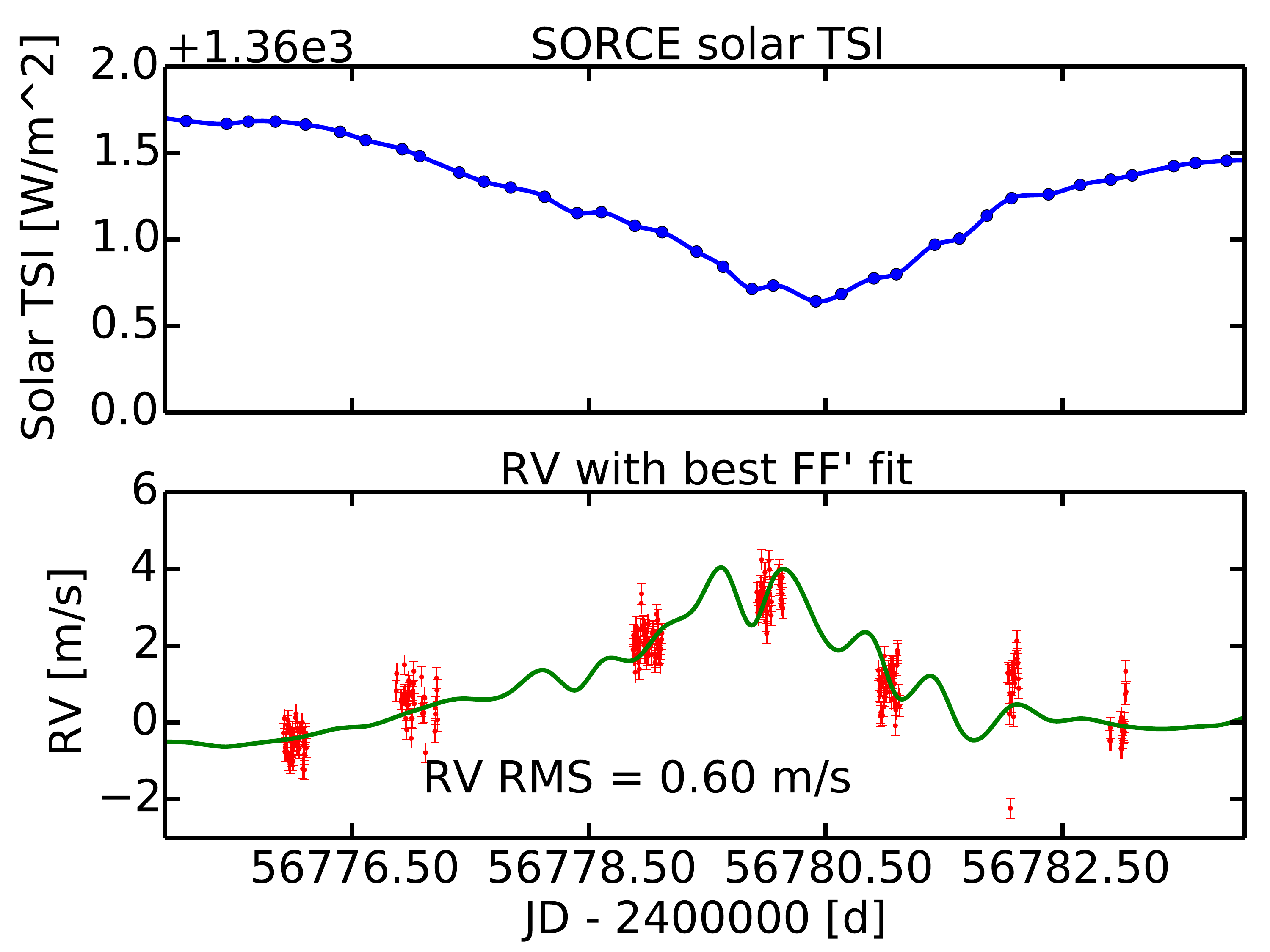}
 \caption{\emph{Top: }Total solar irradiance (TSI) of the Sun as measured by the SORCE satellite. The dots represent the raw SORCE data, and the line is a cubic interpolation. \emph{Bottom: }RVs obtained with the HARPS-N solar telescope. The green curve represent the best FF${}^\prime$ fit to the data using the interpolated SORCE data. The RV rms of the residuals after subtracting the best FF${}^\prime$ model is 60\cms.}
 \label{fig:3-2}
 \end{center}
 \end{figure}

To perform the fit of the RV data using the FF${}^\prime$ method, we used data from the SORCE satellite~\citep{SORCE} that gives the total solar irradiance (TSI), i.e., the disk integrated flux, every 6 hours. We then interpolated the SORCE data to match the RV observations and hypothesized that both parameters $a$ and $b$ are constant on the timescale of the observing run. This is a reasonable argument considering that large spots and plages, which are likely responsible for the majority of the RV variation, are evolving on the timescale of a rotational period, i.e., 25 days \citep[][]{Dumusque-2011b,Howard-2000}. In Figure \ref{fig:3-2} the top panel shows the TSI as a function of time, and the bottom panel represents the RV of our observing run, with the best FF${}^\prime$ fit in green. After removing the best FF${}^\prime$ fit to the RVs, the residuals show an RV rms of 60 \cms for the full observing run. This is an improvement of 55\% compared to the RVs without the FF${}^\prime$ correction. 
 
In the studied dataset, the FF${}^\prime$ method seems to perform efficiently in reducing the scatter induced by stellar noise. However, several
studies have shown that FF${}^\prime$ cannot explain fully the RV variation induced by stellar noise \citep[][]{Haywood-2014,Aigrain-2012}. We therefore 
expect that the FF${}^\prime$ method will not perform as well for future solar RV datasets, and we want to investigate further the use of SDO images and 
dopplergrams in correcting for stellar noises.

\section{Conclusion} \label{sect:4}

First tests of the HARPS-N solar telescope show sub\,-\ms RV precision over a week timescale. More extensive RV measurements with the same quality, combined with the wealth of information available from solar satellites, should allow us to develop a better understanding of the RV variation induced by solar surface inhomogeneities. We expect this will lead to better correction techniques for stellar noise applicable to bright stellar RV targets, and eventually will enable the RV detection of Venus, thus demonstrating the possibility of detecting Earth twins around other stars using Doppler spectroscopy. 
%For stars other than the Sun, it will not be possible to obtain spatially resolved information for the status of the stellar surface. Therefore, we need to use observables that can be obtained directly from spectroscopy and photometry of bright targets when designing those new correction methods. We will explore and develop such methods by observing the Sun with the HARPS-N solar telescope.

\acknowledgments
%{\bf We thanks the anonymous referee for his valuable comments and suggestions that improved the first version of the paper. 

This work was performed with support from the Smithsonian Astrophysical Observatory, The Harvard Origins of Life Initiative, the National Science Foundation and NASA (for support of the astro-comb). X.D. thanks NASA's Transiting Exoplanet Survey Satellite (TESS) mission for partial support via subaward 5710003554 from MIT to SAO. ACC acknowledges support from STFC grant ST/M001296/1 during the course of this work. The research leading to these results received funding from the European Union Seventh Framework Programme (FP7/2007-2013) under grant agreement number 313014 (ETAEARTH). This publication was made possible through the support of a grant from the John Templeton Foundation. The opinions expressed in this publication are those of the authors and do not necessarily reflect the views of the John Templeton Foundation. We are grateful to all technical and scientific collaborators of the TNG telescope and the HARPS-N Consortium that have made this project possible.

\bibliographystyle{apj}
\bibliography{dumusque_bibliography}

\end{document}